\pdfoutput=1

\documentclass[a4paper,10pt]{article}
\usepackage[T1]{fontenc}
\usepackage{lmodern}
\usepackage{lipsum}
\usepackage{pdflscape}
\usepackage{rotating}
\usepackage{caption}
\usepackage{authblk}
\usepackage[top=2cm, bottom=2cm, left=2cm, right=2cm]{geometry}
\usepackage{fancyhdr}
\usepackage{url}
\usepackage{graphicx}
\usepackage{amssymb, gensymb}
\usepackage{amsmath}
\usepackage[authormarkuptext=id]{changes}
\definechangesauthor[name={Drosos}, color=red]{DK}
\usepackage{soul}
\usepackage{tikz}
\usepackage{pgfplots}
\pgfplotsset{compat=1.8}
\usetikzlibrary{patterns}
\usepackage{pgfplotstable, booktabs}  
\definecolor{mycolor1}{RGB}{130,220,202}
\definecolor{mycolor2}{RGB}{79,122,142}  
\definecolor{mycolor3}{RGB}{170,35,3}
\definecolor{mycolor4}{RGB}{207,170,114}
\definecolor{mycolor5}{RGB}{80,135,63}
\definecolor{mycolor6}{RGB}{255,140,190}

\begin{document}

  \title{\sf Magnetic Skyrmions in FePt Square-Based Nanoparticles Around Room-Temperature} 
  
\author[1]{Christos Tyrpenou}
\author[1]{Vasileios D. Stavrou}
\author[1]{Leonidas N. Gergidis\thanks{lgergidi@uoi.gr;\; lgergidis@gmail.com}}
\affil[1]{Department of Materials Science and Engineering, University of Ioannina, 45110 Ioannina, Greece}

\maketitle
\thispagestyle{empty}
\begin{abstract}
  Magnetic skyrmions formed at temperatures around room temperature in square-based parallelepiped magnetic FePt nanoparticles with perpendicular magnetocrystalline anisotropy (MCA) were studied during the magnetization reversal using  micromagnetic simulations. Finite Differences (FD) method  were used for the solution of the Landau-Lifshitz-Gilbert equation. Magnetic configurations exhibiting N\'eel  skyrmionic formations were detected. The magnetic skyrmions can be created in different systems generated by the variation of external field, side length and width of the squared-based parallelepiped magnetic nanoparticles. Micromagnetic configurations revealed a variety of states which include skyrmionic textures with one distinct skyrmion formed and being stable for a range of external fields around room-temperature. The size of the formed N\'eel skyrmion is calculated as a function of the external field, temperature, MCA and nanoparticle's geometrical characteristic lengths which can be adjusted to 
  produce N\'eel type  skyrmions on demand having diameters down to 12 nm. 
  The micromagnetic simulations revealed that stable skyrmions at the  
  temperature range 270 - 330 K can be created for FePt magnetic nanoparticle systems lacking of chiral interactions such as Dzyaloshinskii-Moriya. 
  \end{abstract}

\date{}
\section{Introduction}
\label{sec:intro}
Magnetic skyrmions are considered by many researchers as dynamic 
candidates for next generation high density 
efficient information encoding providing enhanced capabilities 
in magnetic writing and storage \cite{Kiselev:2011,Romming:2013,YuRoom:2016,Sitte:2018,Ezawa:2020}. 
Their small size and  the low current density 
needed for their manipulation and control \cite{Fert2017} pave the way for 
potential engineering applications. 
Over the last decade magnetic skyrmions have been 
under intense theoretical, computational and experimental
investigation \cite{Seidel:2016, Skyrmbook:2017, Buttner:2018, Moutafis:2016, Malik:2018, Sapozhnikov:2018, Fangohr:2018, Schaffer:2019,LeonovNeel:2015}.
It is of significant importance to understand 
how the confined nature of geometry can affect  
skyrmion creation-stabilization \cite{Sampaio:2013,Beg:2015,LeonovPRL:2016} 
and consequently the size and energetics of the skyrmionic states. 

Recently, Fert and his collaborators \cite{Fert:2018} combined concomitant 
magnetic force microscopy and Hall resistivity measurements to 
demonstrate the electrical detection of sub-100 nm skyrmions in a 
multilayered thin film at room-temperature. 
Additionally, Legrand et al. \cite{Legrand:2019} showed that 
room temperature antiferromagnetic skyrmions can be stabilized
in synthetic antiferromagnets (SAFs), in which perpendicular magnetic anisotropy, antiferromagnetic coupling and chiral order can be adjusted concurrently. 
In up to date reports room-temperature magnetic skyrmions in 
ultrathin magnetic multilayer structures of Co/Pd \cite{Pollard:2017} , Pt/
Co/MgO \cite{Boulle:2016} and Co/Ni \cite{Hrabec:2017} were reported. Br\~andao et al. \cite{Brandao:2019} reported on the evidence of 
skyrmions in unpatterned symmetric Pd/Co/Pd multilayers at
room temperature without prior application of neither electric current nor magnetic field.
Husain  et al. \cite{Husain:2019} observed stable skyrmions in unpatterned Ta/Co$_2$ FeAl(CFA)/MgO thin film heterostructures at room temperature in 
remnant state employing magnetic force microscopy showing that these skyrmions consisting of ultrathin ferromagnetic CFA Heusler alloy result from strong interfacial Dzyaloshinskii-Moriya interaction (i-DMI).  Ma et al. \cite{Ma:2019} employed atomistic stochastic Landau-Lifshitz-Gilbert simulations to investigate skyrmions in amorphous ferrimagnetic GdCo revealing that a significant reduction in DMI below that of Pt is sufficient to stabilize ultrasmall skyrmions even in films as thick as 15 nm. 

In the present simulation work, the formation of  magnetic
skyrmions around room-temperature  
in the external field range used for   
the magnetization reversal in a single 
FePt nanoparticle having parallelepiped geometry 
of square base is studied.    
The current research effort was inspired-initiated by the variety of 
skyrmionic magnetic textures  
detected in FePt triangular and reuleaux prismatic 
nanoparticles encapsulating multiple skyrmions 
at 0 K using Finite Elements micromagnetics 
simulations \cite{Stavrou:2016, Gergidis:2019, StavGerg:2019}. 
It was shown that for  magnetocrystalline anisotropy (MCA) values 
between $Ku\mathrm{=200-500 \; kJ/m^3}$ three distinct 
skyrmions are formed and
persist for a range of external fields along with rich  
magnetic structures in the bulk of the FePt nanoparticle 
similar to those observed in DMI stabilized 
systems \cite{Hesjedal:2018}. N\'eel type skyrmionic textures 
were detected on the Reuleaux geometry's bases coexisting 
with Bloch-type textures in the bulk of the FePt nanoelement. It 
was also shown that the size of skyrmions depends linearly  
on the field value $B_{ext}$ and that the slope of the linear curve 
can be controlled by  MCA value \cite{StavGerg:2019}. 
Single layered square-based FePt nanoparticles in the 
absence of chiral interactions like 
Dzyaloshinskii-Moriya (DM) or 
its interfacial analogue (iDM) (present in multi-layered hetero-structures with 
spin-orbit coupling) are investigated. Skyrmionic textures and  formations 
are explored along with the underlying physical 
mechanisms for temperatures around 300 K and for 
different geometrical characteristics of the nanoparticle.     

\section{Micromagnetic modeling}
\label{sec:model}
The Landau-Lifshitz-Gilbert (LLG) equation governs the rate of change of the 
dynamical magnetization field $\mathbf{M}$ and is given by the relation  
\begin{equation}
\frac{d\mathbf{M}}{dt}=\frac{\gamma}{1+\alpha^2}(\mathbf{M} \times \mathbf{B}_{eff})
-\frac{\alpha \gamma}{(1+\alpha^2)|\mathbf{M}|} \mathbf{M} \times (\mathbf{M}\times \mathbf{B}_{eff}).
\end{equation}
The quantity  $\alpha>0$ is a phenomenological dimensionless damping 
constant that depends on the material and $\gamma$ is the electron 
gyromagnetic ratio. The effective field that governs the 
dynamical behavior of the system has contributions from various 
effects that are of very 
different nature and can be expressed as 
$\mathbf{B}_{eff}=\mathbf{B}_{ext}+\mathbf{B}_{exch}+
\mathbf{B}_{anis}+\mathbf{B}_{demag}+\mathbf{B}_{thermal}$.
Respectively, these field
contributions are the external field $\mathbf{B}_{ext}$,
the exchange field $\mathbf{B}_{exch}$, the anisotropy
field $\mathbf{B}_{anis}$, the demagnetizing field $\mathbf{B}_{demag}$ and the thermal field 
$\mathbf{B}_{thermal}$.

In particular, the thermal field $\mathbf{B}_{thermal}$ which 
incorporates implicitly the temperature 
can be expressed by 
$\mathbf{B}_{thermal}(t)=\mathbf{n}(t) \sqrt{\frac{2\mu_{0} \alpha k_BT}{B_{sat}\gamma \Delta V \Delta t}}$ according to 
Brown \cite{Brown:1963}. In thermal field expression $\alpha$ is 
the aforementioned damping constant, $\mu_0$ is the 
vacuum permeability constant,  $k_B$ is the Boltzmann constant, $T$ is the 
temperature (throughout manuscript  slightly slanted $T$  
is used for temperature) , $B_{sat}$ the saturation magnetization expressed in Tesla (T) 
, $\Delta V$ the 
cell volume, $\Delta t$ the time step and $\mathbf{n}$ a random vector 
generated from a standard normal distribution whose 
value is changed every time step. 
For the solution of the LLG equation  
micromagnetic Finite Differences (FD) calculations have been 
conducted  using  Mumax3 Finite Differences software \cite{Mumax:2014, Mumax:2018}. The dimensionless damping constant $\alpha$ was set to 1 in order to achieve fast damping and
reach convergence quickly as we are interested in static magnetization configurations. 
The time step used for the integration of the LLG equation 
was equal to  $\Delta t=\mathrm{1 \; fs}$. 
 
The square-based nanoparticle having side 
length $\textsl{\textrm{a}}=\mathrm{150 \; nm}$  
and width $w=\mathrm{36 \; nm}$ will be considered 
as the reference nanoparticle  and will be used for the 
multi-parametric investigation conducted in 
the present work. It is chosen since it encapsulates 
the most stable skyrmion with respect 
to the magnetic field range and skyrmion type 
as it will be made clear in the sequel.
The particular width value $w=36$ nm is selected following previous 
studies  \cite{Stavrou:2016,Gergidis:2019,StavGerg:2019} which  
focus on FePt triangular and reuleaux prismatic 
nanoelements. In addition, the  36 nm thickness of nanoparticle matches that reported in~\cite{Markou:2013}. 
For the objectives of the multi-parametric study additional 
magnetic samples-nanoparticles with parallelepiped geometry 
of square base with varying side length $\textsl{\textrm{a}}$ 
from 40 to 180 nm and  $w$ ranging from 6 to 36 nm were 
also used. The following frame of reference axes assignment 
convention was used: $x,y$  along the 
square's edges, and $z$ perpendicular to the nanoparticle's  square base. 
The  mesh used for the discrete representation of the rectangular parallelepiped nanoparticle 
 under study was a regular 3D mesh with characteristic discretization lengths $\Delta x= \Delta y$ =2 nm, $\Delta z$=1 nm in $x,y,z$-directions 
respectively. The lengths  $\Delta x,\Delta y,\Delta z$ used for the discretization of the rectangular domain under investigation were lower than the exchange length $l_{ex}=\sqrt{\frac{2A}{\mu_0M_s^2}} \approx 3.5 $ nm 
for the FePt magnetic material. 

The material parameters used in this study have been   
reduced accordingly in order to follow their 
temperature variation. Since the exact 
temperature dependence is not exactly known the reduction of the 
anisotropy parameter and micromagnetic exchange  takes place through 
the approximate relations $K(T) \sim [m_e(T)]^3$, 
$A_{exch}(T) \sim [m_e(T)]^2$ \cite{Atxitia2007,Callen1966}. The spontaneous equilibrium magnetization $m_e(T)$ is taken from the atomistic FePt model 
presented in   \cite{Kazantseva:2008}. The dependence of saturation magnetization on temperature was modeled using the findings of Okamoto et al. \cite{Okamoto:2002}. 
The MCA was oriented perpendicular to the nanoparticle's 
square base and  the MCA constant 
was varied. The  $Ku\mathrm {=250 \; kJ/m^3}$ in our previous 
micromagnetic numerical endeavors \cite{Gergidis:2019,StavGerg:2019} at 0 K 
was capable of creating interesting skyrmionic textures in triangular and Reuleaux based FePt nanoelements.

The Finite Differences (FD) micromagnetic simulations were conducted 
on Nvidia GTX 1080 GPU.  The magnetization curves for every production 
run were investigated by applying external magnetic 
fields $B_{ext}$ with fixed orientation 
running parallel to $z$-direction (the normal to the 
nanoparticle's square base). 
The range values of  $B_{ext}$ were 
 +1 T (maximum) and -1 T (minimum) having an external 
 magnetic field step of $\delta B_{ext}$ = 0.01 T for 
 the actual magnetic reversal process.
%
\section{Results}
\label{sec:Results}
\subsection{Skyrmion formation and stabilization for the reference nanoparticle}
Skyrmion formation and stabilization as a function of the external 
field, the geometric characteristics of the nanoparticle
and of the temperature can be quantitatively characterized 
by the calculation of the topological invariant $S$, widely 
known as skyrmion or winding number  \cite{Skyrme:1962,Skyrmbook:2017,Palotas:2018}.  
The skyrmion number $S$ is computed using the relation  
$S= \frac{1}{4\pi} \int_A   \mathbf{m} \cdot (\frac{\partial\mathbf{m}}{\partial x} \times \frac{\partial \mathbf{m}}{\partial	y} ) dA$. The quantity $\bf{m}$ is 
the unit vector of the local magnetization defined as
$\mathbf{m}=\mathbf{M}/M_s$ with $\mathbf{M}$ being the magnetization and $M_s$ the saturation magnetization. Magnetization $\mathbf{M}$  is 
provided by the FD numerical solution of the LLG equation. 
The skyrmion number $S$ is a physical and topological
quantity that  measures how many times $\mathbf{m}$ wraps the unit sphere \cite{Skyrmbook:2017, Yoo:2014, Fangohr:2015}. 
Surface $A$ is the surface
domain  of integration and corresponds to the square base 
of the FePt nanoparticles under investigation. 

\begin{figure}[!h]
	\centering
	\includegraphics[totalheight=4.6in,angle=0]{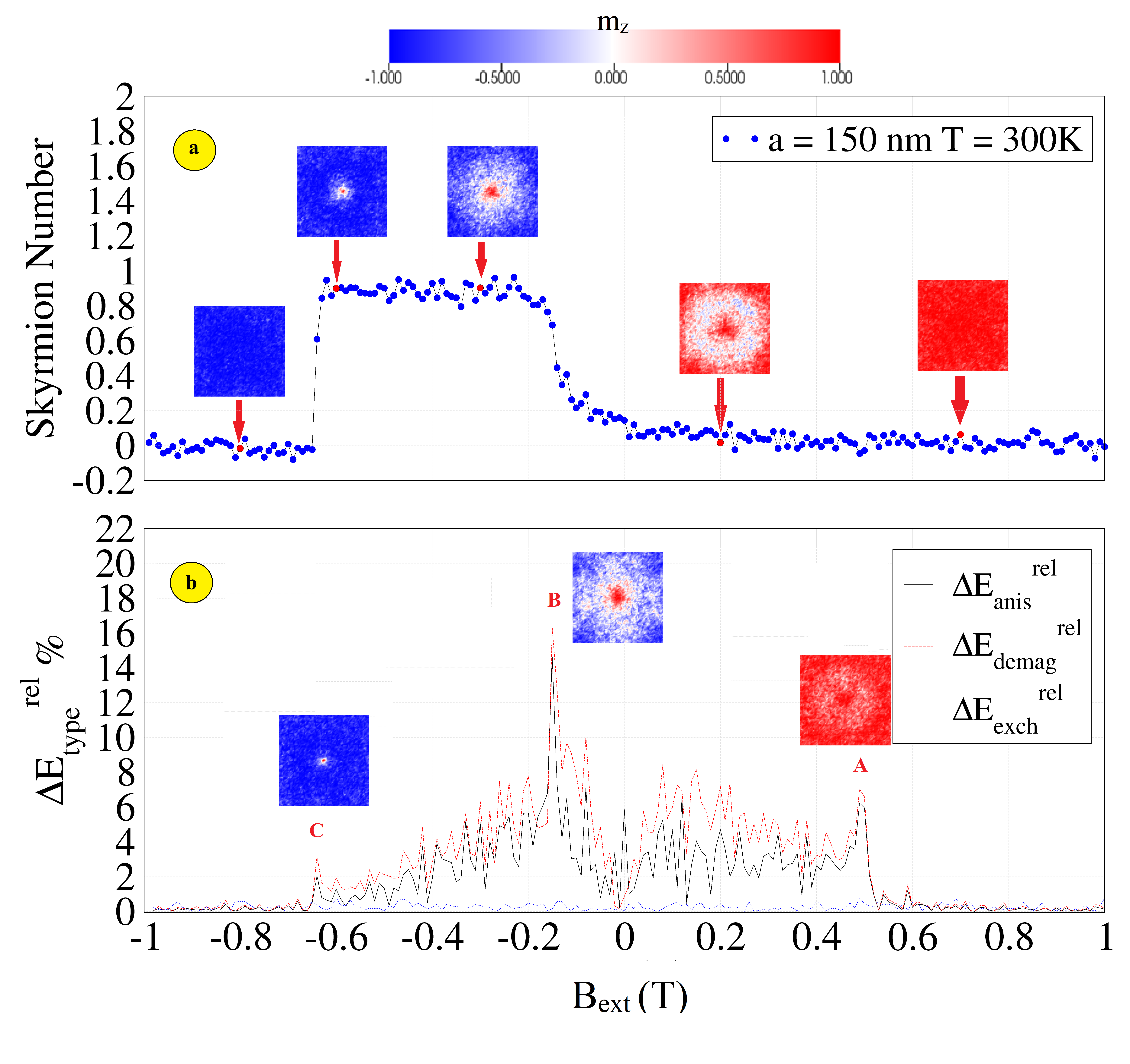} 
	\caption{Skyrmion number $S$ as a function of $B_{ext}$ for the reference nanoparticle at 300 K (panel a). The pseudo color which is shown in the depicted micromagnetic configurations refers to the $z$-component of magnetization ($m_z$). 
		Relative energy differences as a function of the external field 
		(panel b).}
	\label{fig:S150}
\end{figure}

The skyrmion number $S$ has been computed during the magnetization 
reversal process as a function of $B_{ext}$ and is shown in 
\textbf{Fig. \ref{fig:S150}a} for the reference 
nanoparticle  at $T=$300 K and 
for $Ku$=130.4 kJ/m$^3$. As the external field decreases 
the magnetic system departs from 
saturation.The  magnetization reversal process 
is closely followed by the external field step 
of $\delta B_{ext}=$ 0.01 T. At first glance for fields down 
to 0.2 T skyrmion number attains very low values fluctuating around zero.  Below 0.2 T the skyrmion number remains fluctuating but attains only positive values indicative for possible starting generation process 
for a specific micromagnetic configuration. The decrease of 
the field below  0 T activates a gradual increase of the 
skyrmion number from $S=0.1$ to $S=0.3$ 
when the external field reaches the  $B_{ext}=-0.1$ T value.  
Infinitesimal decrease of the magnetic field 
below  $B_{ext}\mathrm{=-0.1 \; T}$    
triggers a jump-like discontinuity on skyrmion number. The value of $S$ 
abruptly increases from $S \approx 0.3$ to $S\approx 1$. Further 
decrease of the external field does not affect the skyrmion number which develops an extended plateau region with $S\approx 1$ for external field values down to 
$B_{ext}\mathrm{=-0.63 \; T}$.     
For magnetic field  value below $B_{ext}\mathrm{=-0.63 \; T}$ a new 
abrupt jump-like reduction of skyrmion number $S$ 
signals the skyrmion annihilation and 
the finalization the magnetization reversal process. 
It is worth mentioning that the existence of thermal field is reflected on the oscillatory behavior of $S(B_{ext})$. These low amplitude oscillations also captured by the micromagnetic configurations visualizations are indicative of the noisy character caused by the thermal field.

The numerical solution of the LLG equation allows the direct 
representation and inspection 
of the micromagnetic configurations. These configurations are also shown  
in \textbf{Fig. \ref{fig:S150}a} for representative 
values of the external field. It should be noted that the pseudo color used 
for the micromagnetic configurations refers to $z$-component 
of magnetization ($m_z$).
At the initial states of the  
reversal process the magnetization vectors  
are aligned parallel to the external field (red colored square in \textbf{Fig. \ref{fig:S150}a}). 
The micromagnetic configurations for fields $B_{ext}\in [-0.14, 0.2]$ T have a donut-like shape where the core magnetizations point 
upwards ($+z$)  and the magnetizations on the distinct 
peripheral (rim) domain are tilted having $m_z=0$ (white color domains in  \textbf{Figure \ref{fig:S150}a}) or reversed ($m_z<0$) 
(light blue color spots also in \textbf{Figure \ref{fig:S150}a}). 
The aforementioned magnetization configuration is a magnetic 
skyrmionium which is a non-topological soliton which has a donut-like out 
of plane spin texture in magnetic nanoparticles and thin films. 
It can be viewed as a coalition of two magnetic skyrmions with opposite skyrmion numbers giving a zero total ($S_{skyrmionium}=0$) \cite{ZhangX:2016,Donutwo:2018}.  The skyrmionium is  considered as a distinct skyrmionic state with  particular characteristics \cite{ZhangX:2016} heralds the abrupt transition to a clear 
skyrmionic state ($S=1$). 
The formed skyrmion which has a perfect 
circular shape is a N\'eel-type  skyrmion and 
as mentioned earlier remains stable 
for a significant  external field range.  In the one 
skyrmion plateau region as the field 
value decreases 
from $B_{ext}\mathrm{=-0.16 \; T}$ to $B_{ext}\mathrm{=-0.63 \; T}$ the actual diameter of the skyrmion also decreases as it can be seen from 
the representative micromagnetic configurations 
shown in \textbf{Fig. \ref{fig:S150}a}. The magnetization reversal process is complete when all magnetization vectors are aligned parallel to $z-$direction (blue colored square in \textbf{Fig. \ref{fig:S150}a}). 

The jump discontinuities of skyrmion number 
as the external field decreases were evident 
in \textbf{Fig. \ref{fig:S150}a}, albeit the detailed mechanism behind this behavior is not clear.  The complex phenomena related to skyrmion formation 
 as well as with the skyrmion number discontinuities could be associated with the rich energetic environment having contributions 
from demagnetization $E_{demag}$, exchange $E_{exch}$ and anisotropy  $E_{anis}$  energies. In order to shed light on the skyrmion formation the  individual energetic contributions \cite{Dai:2013} were quantified during the magnetization reversal process by computing the 
absolute relative energy difference 
$\Delta E_{type}^{rel,i}=|\frac{E_{type}^{i+1}-E_{type}^{i}}{E_{type}^{i}}|\times
100(\%)$ (where $type$ stands for $anis,exch,demag$) 
between the consecutive ($i$ and $i+1$) external magnetic
field values $B_{ext}^{i},B_{ext}^{i+1}$ with ($i=0 , 199$).   
The values of the relative differences 
of anisotropy, demagnetization 
and exchange energies are also shown in \textbf{Fig. \ref{fig:S150}b } as functions 
of $B_{ext}$ for  $Ku\mathrm{=130.4 \; kJ/m^3}$.

Initially, the relative energy differences have very low 
fluctuating values  as external field decreases down to 
$B_{ext}=\mathrm{0.5 \; T}$  where a 
discontinuity can be observed for all relative anisotropy and demagnetization energies and 
clearly can be associated 
with the creation process of the aforementioned skyrmionium texture. 
The relative difference values  
are 6\%, 6.5\%  for $\Delta E_{anis}^{rel}, \Delta E_{demag}^{rel}$, 
respectively. 
Gradual decrease of the external field is followed by 
demagnetization and  exchange energy fluctuations for field values 
down to -0.14 T. Further decrease of the 
 field triggers the skyrmion formation followed by abrupt jump discontinuities in  $\Delta E_{anis}^{rel}, \Delta E_{demag}^{rel}$ 
 energies at -0.14  T. The formed skyrmion ($S\approx 1$) at -0.14 T remains stable as the field further decreases with $\Delta E_{anis}^{rel}, \Delta E_{demag}^{rel}$ retaining their fluctuating 
 character showing a clear  tendency to attain lower values. At the field value of -0.63 T a new jump discontinuity but with significantly reduced jump amplitude is evident  on the relative energy differences shown in  \textbf{Fig. \ref{fig:S150}b} which signals the skyrmion annihilation.  
 The relative difference values at the skyrmion annihilation discontinuity are 2\%, 3\%  for anisotropy, demagnetization, respectively. It is worth noting that   $\Delta E_{exch}^{rel}$ (< 0.7\%) remains practically constant with weak fluctuations during the magnetization reversal process while the demagnetization and anisotropy energies play the crucial role to skyrmion formation. 

\subsection{Skyrmion dependence on nanoparticles's width}
For nanoparticles having $\textsl{\textrm{a}}=\mathrm{150 \; nm}$ 
individual micromagnetic simulations have been conducted at different 
width ($w$) values in order to investigate the effect 
of the nanoparticle's width on 
the actual skyrmion formation during the magnetization reversal process 
at the temperature of 300 K. 
\begin{figure}[!t]
	\centering
	\includegraphics[totalheight=4.0in,angle=0]{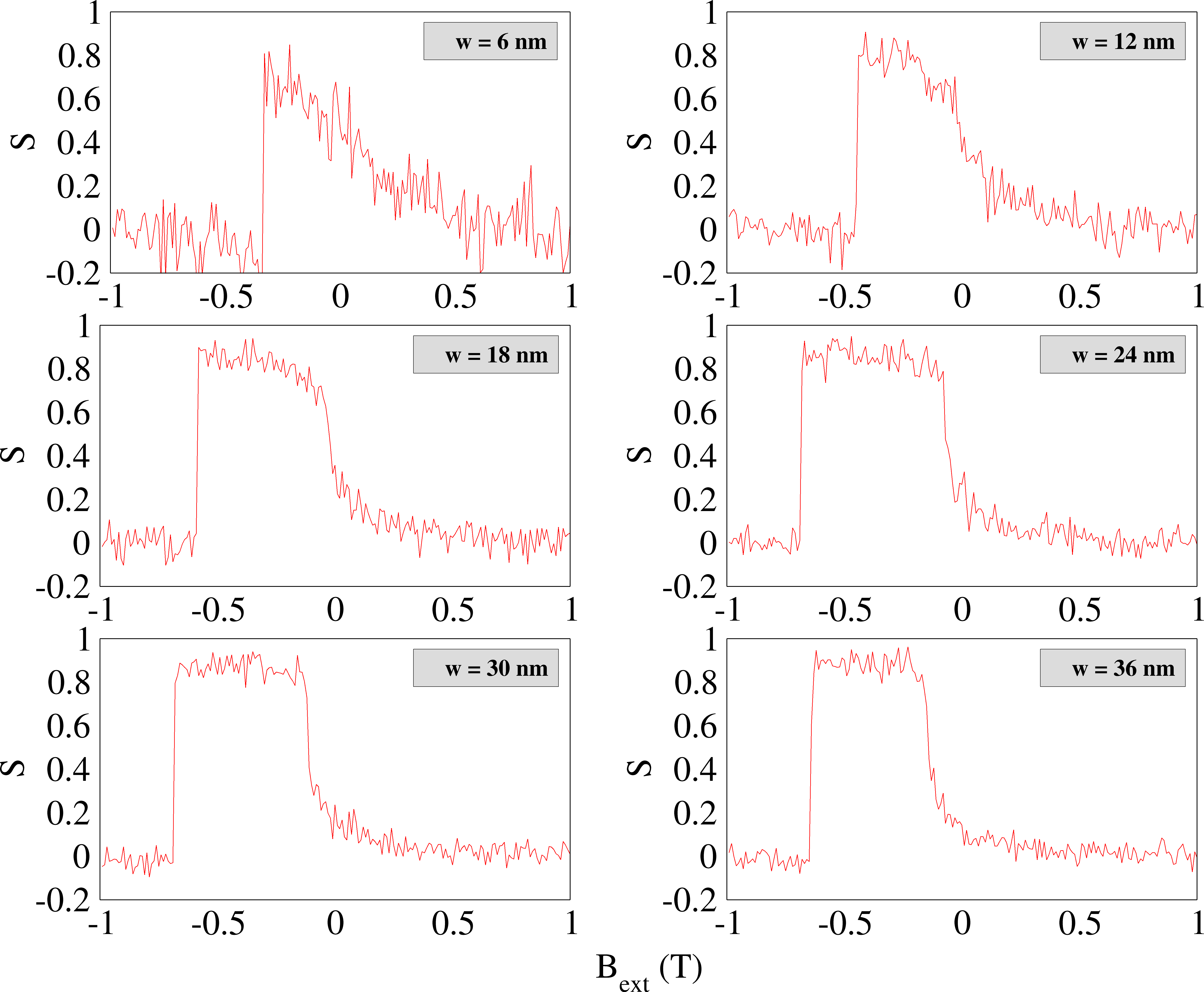} 
	\caption{Skyrmion number $S$ as a function of $B_{ext}$ for  side value $\textsl{\textrm{a}}=\mathrm{150\; nm}$ at 300K for geometries having different widths $w=6, 12, 18, 24, 30, 36$ nm.}
	\label{fig:varw150nm}
\end{figure}
The skyrmion number $S$ as a function of $B_{ext}$ is shown for width values 
$w$=6 - 36 nm  in \textbf{Fig. \ref{fig:varw150nm}}. 
Skyrmion can be produced in substantial ranges even for 
widths down to 12 nm. 
The present FD micromagnetic simulations reveal that the 12 nm width 
for nanoparticles having $\textsl{\textrm{a}}=\mathrm{150 \; nm}$ is critical for the creation of skyrmions. Nanoparticles with width values 
below 12 nm cannot generate complete skyrmions.  The $S(B_{ext})$ line shapes for the nanoparticle with  $w=\mathrm{6 \; nm}$ 
substantially deviate from the respective line shapes for higher widths. 
A smooth increase (still fluctuating) of the skyrmion 
number as a function of the reducing external magnetic field during the 
skyrmion's creation process is evident. The 
nanoparticles with $w=\mathrm{12-36 \; nm}$ expose a 
discontinuous jump-like character 
for external fields around 0 T. The higher the width the higher the jump-like discontinuity. For all widths the skyrmion annihilation 
which occurs at  different field values is an abrupt process. The skyrmion number reduces from $S\approx 1$ to $S\approx0$ with no 
intermediate transition magnetic states.  

\subsection{Skyrmion dependence on MCA}
\begin{figure}[!h]
	\centering
	\includegraphics[totalheight=3.0in,angle=0]{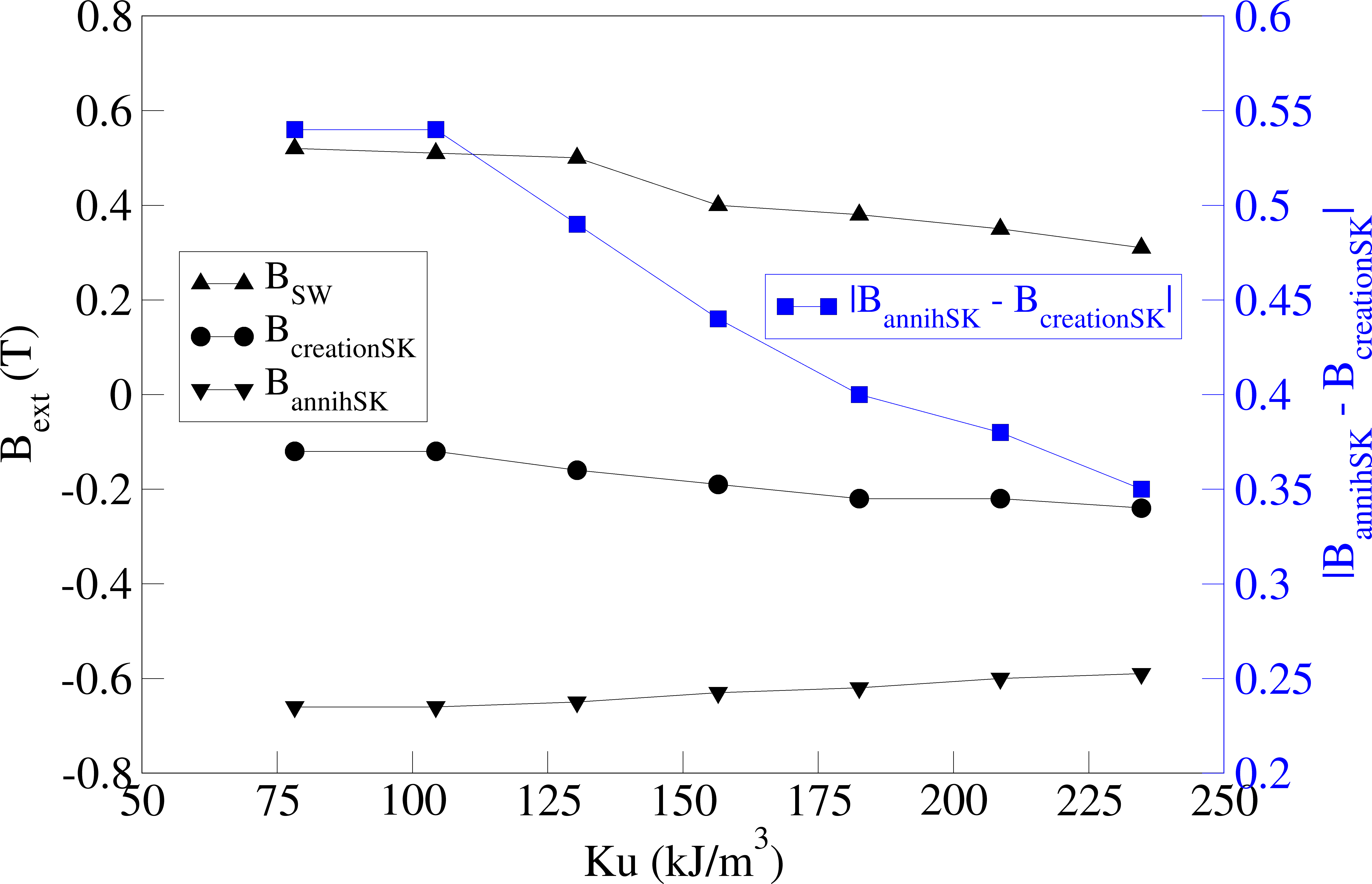} 
	\caption{External fields relative to the formation ($B_{creationSK}$) and annihilation ($B_{annihSK}$) as a function of MCA value ($Ku=\mathrm{78.2-234.7 \; kJ/m^3}$) for the 
		reference nanoparticle at 300 K.}
	\label{fig:anisotropies}
\end{figure}

For the  reference nanoparticle ($\textsl{\textrm{a}}=\mathrm{150 \; nm}$, 
$w=\mathrm{36 \; nm}$) simulations  have been conducted 
with varying MCA values $Ku=\mathrm{78.2-234.7 \; kJ/m^3}$ 
while its orientation is set 
parallel to external field and $z$-direction. It is interesting the fact that MCA 
mainly affects the magnetic field range of stabilization of 
the skyrmion formed and not the skyrmionic state of $S=+1$. 
Two external fields during the creation ($B_{creationSK}$) and 
the annihilation ($B_{annihSK}$) of skyrmion were recorded and 
presented in \textbf{Fig. \ref{fig:anisotropies}}. The $B_{creationSK}$ values of the magnetic field follow a rather weak decrease from -0.12 T 
to -0.25 T with the gradual increase of the MCA from  $Ku=\mathrm{78.2 \; kJ/m^3}$ to $Ku=$234.7  kJ/m$^3$. 
The skyrmion annihilation field value of $B_{annihSK}$ exposes a 
weak but gradual increase from -0.7 T to -0.6 T as the 
MCA value increases. 

The external magnetic 
field skyrmion's stabilization range $|B_{annihSK}-B_{creationSK}|$ is also monitored in \textbf{Fig. \ref{fig:anisotropies}} exposing a  monotonic decrease as MCA value increases from $Ku$=104.3 and  234.7 kJ/m$^3$. 
The switching field which is related to initiation of the reversal 
process is affected by the MCA value showing a clear reduction from 0.5 to 
0.3 T with the increase of $Ku$.  
\subsection{Skyrmionic states}
\begin{figure}[!b]
	\centering
	\includegraphics[totalheight=2.2in,angle=0]{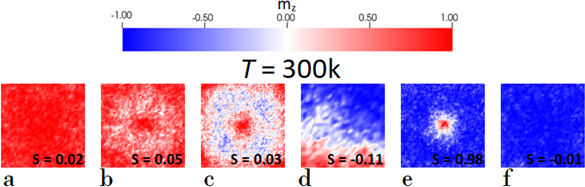} 
	\caption{Different states detected for the square base FePt nanoparticle (reference geometry). 
		The states are: (a, f) uniform, (b, c) skyrmionium, 
		(d) domain wall, (e)skyrmion with $S=+1$. MCA value was $Ku\mathrm {=130.4 \; kJ/m^3}$}
	\label{fig:states}
\end{figure}

The topological invariant $S$ reported along with the 
actual magnetic configurations can provide valuable 
qualitative information toward the detection and characterization of 
skyrmionic states. It is very interesting 
the fact that a variety of micromagnetic skyrmionic states 
emerges for the square-based FePt nanoparticles. 
The most commonly detected micromagnetic states 
are presented along with the respective skyrmion 
number $S$  in \textbf{Fig. \ref{fig:states}} and can be assigned to the following categories: 
\\
\textbf{a, f.} \textit{Uniform states}: States where the magnetization is uniform and the actual magnetization vectors point all upwards (red color) or all downwards (blue color)
\cite{Chui:2015,Fangohr:2018}.
\\
\textbf{b, c.} \textit{Skyrmionium}: A magnetic skyrmionium is a non-topological soliton of a doughnut shape \cite{ZhangX:2016,Donutwo:2018}. 
\\
\textbf{d.} \textit{Domain wall}:  A domain wall is a gradual reorientation of individual magnetic moments across a finite distance undergoing an angular displacement of 90$^o$ or 180 $^o$ \cite{coey:2010}.  
\\
\textbf{e.} \textit{Skyrmion with $S = 1$}: State where one skyrmion is formed \cite{Palotas:2018, Gergidis:2019}.
\\
%

\begin{sidewaysfigure}
	\includegraphics[totalheight=5.2in,angle=0]{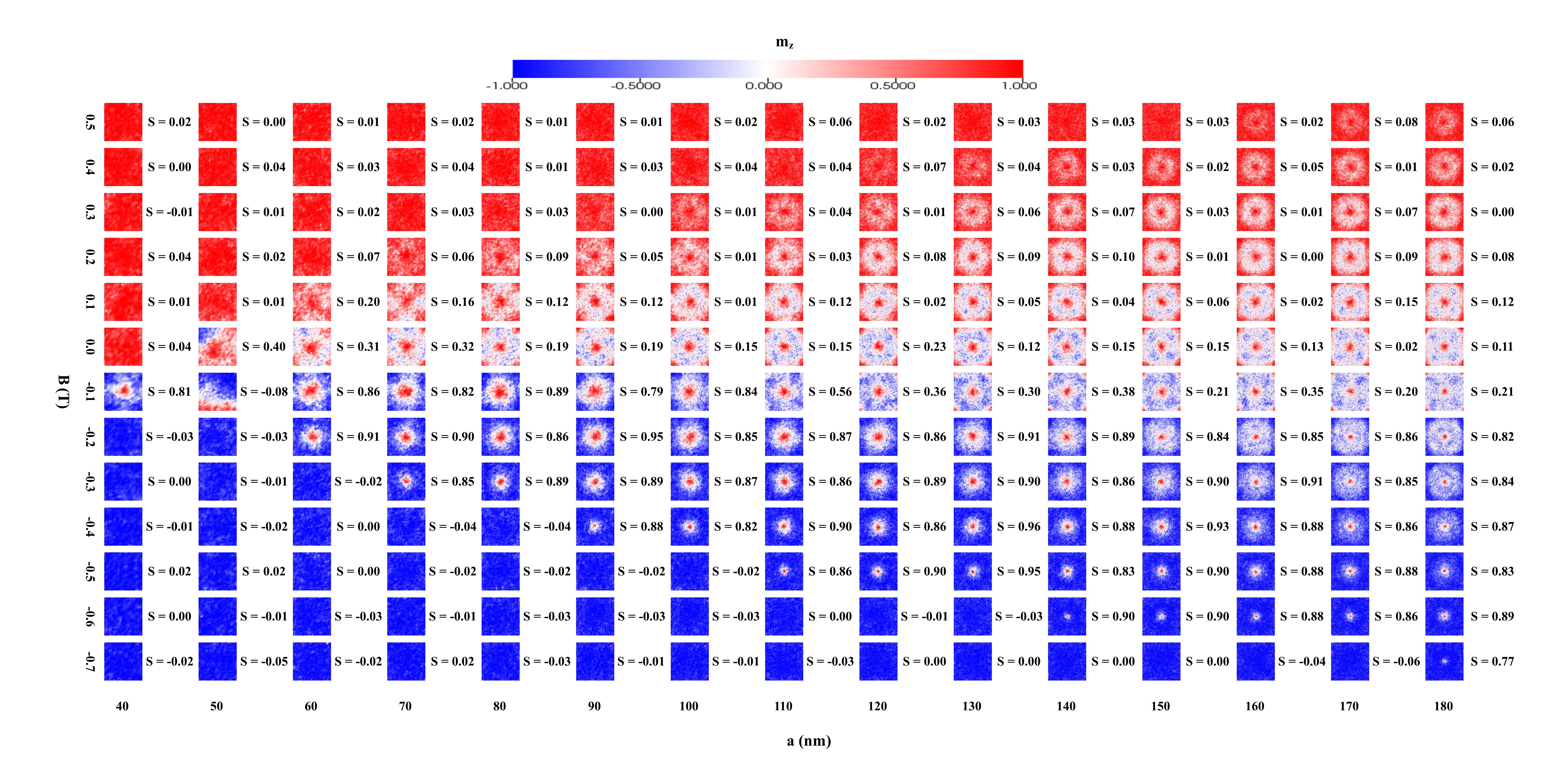} 
	\caption{Micromagnetic states revealed at 300 K for the FePt 
		nanoparticle. The color bar refers to the $z$-component of magnetization ($m_z$).}
	\label{fig:totstates}
\end{sidewaysfigure}

From the results presented and the conclusions drawn 
so far it is clear that skyrmionic 
textures can be created and stabilized by adjusting 
the width of the nanoparticle, the MCA value and the 
temperature of the magnetic system. 
In addition, the skyrmion number $S$ has been computed during the 
magnetization reversal for  FePt nanoparticles with variable 
side length. The \textbf{Fig. \ref{fig:totstates}} depicts for the temperature of 300 K, $Ku\mathrm {=130.4 \; kJ/m^3}$ and $w=36$ nm a skyrmionic state diagram 
constructed by the different values of external field $B_{ext}$ and of 
the square side length $\textsl{\textrm{a}}$. 
Nanoparticles with side length $\textsl{\textrm{a}}\le$ 50 nm are not capable of hosting stable skyrmionic entities irrespective of the applied external field as 
can be seen in \textbf{Fig. \ref{fig:totstates}}. The restricted surface 
area affects also the precursor skyrmionium state which is absent in the aforementioned nanoparticles. The precursor skyrmionium states for positive external field values start to appear assisting the development of 
N\'eel skyrmions for square bases 
with $\textsl{\textrm{a}} \ge 60$ nm. The skyrmionium states are present and their development is intimately related to the side length $\textsl{\textrm{a}}$. The higher the side length the higher 
the value of the characteristic positive external field value in which they  appear. It is very interesting the fact that for the square based 
parallelepiped geometry the magnetic skyrmions are created 
for $B_{ext}<0$ values and in particular for -0.1 T.  
The detected skyrmions are N\'eel skyrmions having skyrmion number values  in the [0.79,0.96] interval. 

Magnetic nanoparticles that promote not only the creation but 
also the  stabilization of skyrmions in a considerable 
range of external fields are the nanoparticles with square side length $\textsl{\textrm{a}} \ge \mathrm{90 \; nm}$.  
For instance the skyrmions created on the nanoparticles with $\textsl{\textrm{a}}= \mathrm{90, 100 \; nm}$ can be persistent for 
the field [-0.4,-0.1] T range while the $\textsl{\textrm{a}}= \mathrm{100, 120, 130 \; nm}$  and $\textsl{\textrm{a}}= \mathrm{140-180 \; nm}$ the annihilation field is in the vicinity of -0.5 and -0.6 T, respectively. 
At this point it should be noted that the micromagnetic states 
represented in  \textbf{Fig. \ref{fig:totstates}} reveal  
a significant symmetry with respect to -0.1 T magnetic field. In the majority of the cases the skyrmionic textures of well developed skyrmioniums  at positive fields have their mirror -0.1 T axis skyrmion analogue created at negative fields.   
%
\subsection{Nanoparticle's internal magnetic structure}
\begin{figure}[!h]
	\centering 
	\includegraphics[totalheight=3.0in,angle=0]{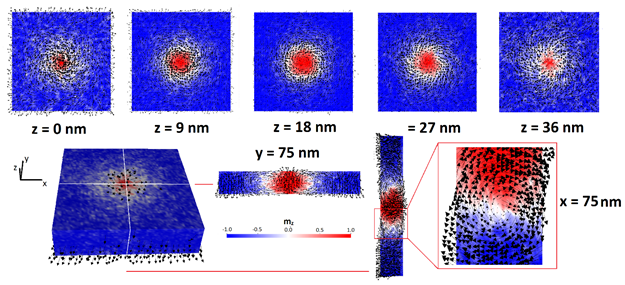} 
	\caption{Micromagnetic configurations of a N\'eel skyrmion at different $z$-levels ($z=$0, 9, 18, 27, 36 nm) for $B_{ext}=-0.3T$ and for the reference nanoparticle at 300K. In addition,  $xz$  and $yz$-cross sections sliced at $y=75, x=75 \mathrm{nm}$ respectively are depicted for the square-base parallelepiped geometry of the magnetic nanoparticle. In particular, a magnified domain of the $yz$-cross section is also shown. The color bar represents the $z$-component of the magnetization ($m_z$).}
	\label{fig:skyrmionslice}
\end{figure}

Finite Difference micromagnetic simulations can provide rich and detailed 
information about the magnetization behavior on the surface and in the bulk 
of nanoparticle. The detected 
skyrmions on the surface are inevitably related 
to the magnetization formations on the internal domain of the nanoparticle.  
The existence of magnetic structure in the internal domain 
of the magnetic nanoparticle  
can be revealed by monitoring the magnetization 
vectors $\mathbf{M}$ for grid points located at 
different $z$-levels ($xy$-cross section) and for $yz$, $xz$ - cross sections.  
These cross sections were chosen for micromagnetic systems hosting 
one N\'eel skyrmion in order to describe the qualitative characteristics 
of the actual skyrmionic texture. N\'eel skyrmion is shown in \textbf{Fig. \ref{fig:skyrmionslice}} for the reference geometry and for $B_{ext}=\mathrm{-0.3}$ T and  MCA value $Ku=\mathrm{130.4 \; kJ/m^3}$. 

At first glance the presence of magnetic structure is evident for 
the N\'eel skyrmion (\textbf{Fig. \ref{fig:skyrmionslice}}) in 
all cross sections visualized.  In particular, $xy$-cross sections  
at $z=\mathrm{0, 9, 18, 27, 36 \; nm}$ 
reveal skyrmionic configurations appearing in each square $z$-level 
and are  presented in  \textbf{Fig. \ref{fig:skyrmionslice}}. 
Interesting is the fact that  
micromagnetic configurations in the non-surface $z$-values 
expose different magnetic characteristics with respect 
to skyrmions observed on the square bases of the FePt nanoparticle. 
On the top or bottom surface N\'eel skyrmions 
are formed while in the proximity of central bulk region 
magnetization vectors deviate 
($z=9,27$ nm of \textbf{Fig. \ref{fig:skyrmionslice}})
from the N\'eel skyrmion pattern adopting a 
more Bloch-type skyrmionic magnetic  
texture at $z=18$ nm with vortex type circulating 
magnetizations. The size of the skyrmionic regions 
at different $z$-levels is different with the skyrmion diameter 
being increased from the surface to the bulk domain of the nanoparticle. 
The higher diameter can be observed for $z=18$ nm.  

Magnetization configurations represented by the magnetization 
vector on the nodal points at $z$=0 nm 
and $z$=36 nm are different as can be observed  
in \textbf{Fig. \ref{fig:skyrmionslice}}. 
It should be pointed out that although  
the two aforementioned $z$-levels where the magnetization 
configurations hosting skyrmions look different
they are topologically equivalent. This can be justified 
by the fact that the calculated integrals of the 
topological density on the two  
square base areas give the same $S$. 

The present simulation results support the depth dependence of helicity which is an interesting result that has been recently observed experimentally by tomography in systems with bulk skyrmions reported 
by Zhang et al. \cite{Hesjedal:2018}. It should be pointed out 
that although the physics  behind skyrmion formation in this system may differ the depth dependence is obviously related to surface effects in both systems. In the present case 
where skyrmions are created in thin nanoparticles the depth dependence of the demagnetizing field (which has increased $z$-component near the surfaces \cite{Tang:2005}) can force the vortex to acquire a N\'eel character at the surfaces while maintaining its typical chiral nature at the bulk. The skyrmionic structures observed here depend on the demagnetizing effects which are proportional to the thickness $w$ (as the easy axis remains in-plane) 
and vanish for low thickness values ($<$6 nm) as it was shown from the presentation 
of the current simulation results and discussion. Inevitably, the rich skyrmionic textures and physical phenomena related to the depth dependence of helicity  are being suppressed as the thickness becomes comparable to the exchange length ($l_{ex}$). 

At $xz$-cross section (sliced at $y$=75 nm) which hosts the N\'eel skyrmion 
the magnetization vectors develop three distinct regions along 
the $x$-direction as exposed in \textbf{Fig. \ref{fig:skyrmionslice}} with 
two clear vortex-like magnetization circulations on the interfaces between the regions.  In the regions (blue-shaded) located at the right and left corners 
of $xz$-cross section magnetizations are pointing  downwards with the 
dimensionless magnetization $z$-component 
attaining values $m_z$=-1.  Moving away form the corners toward the 
skyrmionic bulk regions the tilting of magnetizations starts producing 
the aforementioned vortex formations at the interfaces.  
At the central bulk region 
the magnetizations start to align parallel to $z$-direction 
pointing upwards ($m_z$=+1). 

The $yz$-cross section is also shown for 
the N\'eel in \textbf{Fig. \ref{fig:skyrmionslice}}. 
A region hosting magnetization circulations is located 
at the center of the $yz$-cross section. Distinct regions hosting parallel magnetizations pointing downwards (blue-shaded region with $m_z$=-1) and upwards  (red-shaded region with $m_z$=+1) are evident. Their  position follows an alternating fashion blue-red-blue 
with the interface regions (white region having $m_z$=0) hosting the center of vortex formations.

\subsection{Skyrmion's size-diameter}
The isolated N\'eel skyrmions that have been 
generated on the FePt nanoparticle of square 
parallelepiped geometries during the reversal process have 
circular geometry with varying diameter $d_{sk}$. Their size 
can be controlled by the magnitude of external field as reported in the recent literature \cite{Wang:2018} and  shown by Finite Element micromagnetics simulations for FePt nanoparticles having reuleaux geometries at 0 K \cite{StavGerg:2019}. The skyrmion's size  dependence on the external field and on the square base 
side length $\textsl{\textrm{a}}$ is computed and quantified. 
\begin{figure}[!h]
	\centering
	\includegraphics[totalheight=4.4in,angle=0]{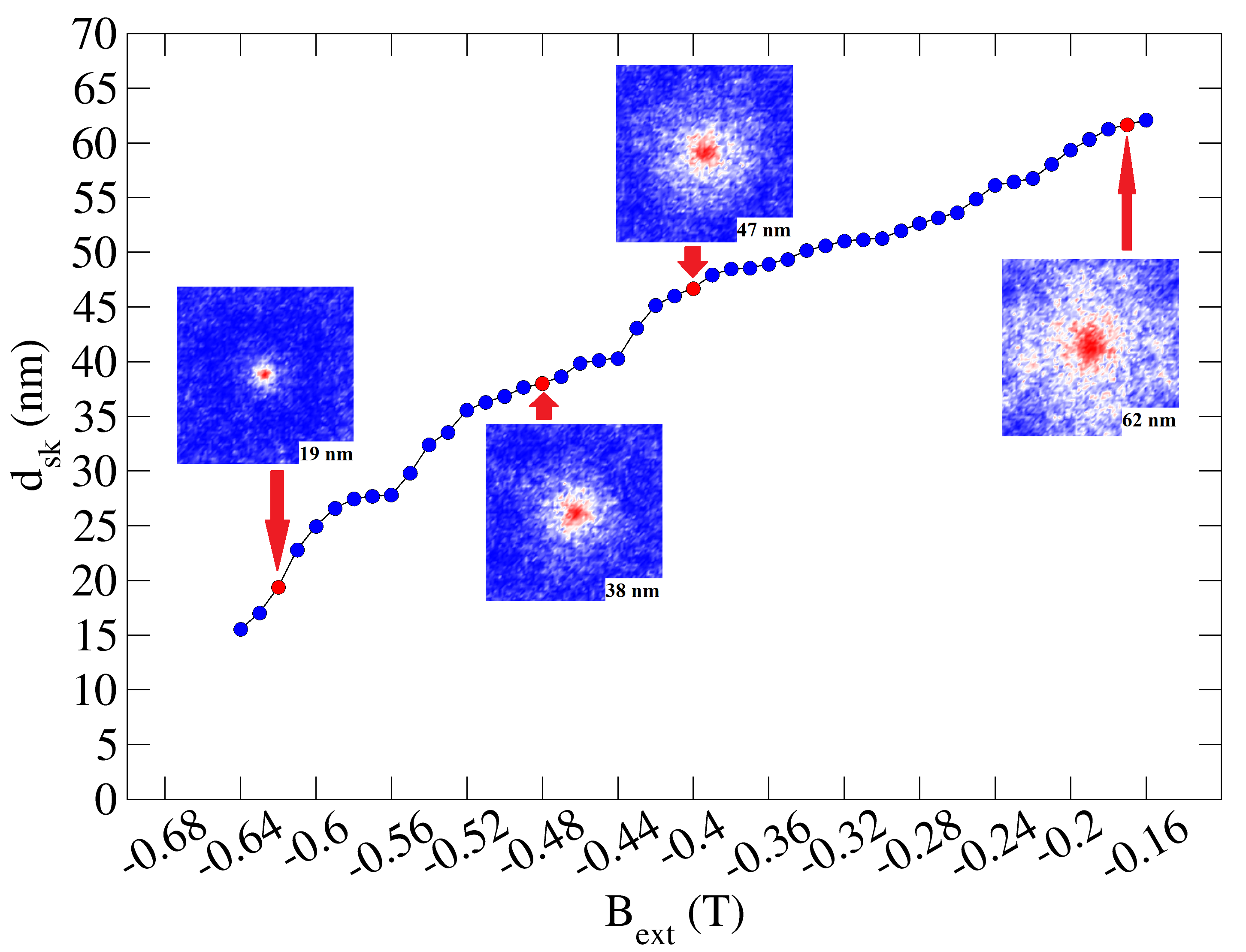}
	\caption{Skyrmion's  diameters $d_{sk}$ at  $T$=300 K as a function of the applied external field $B_{ext}$ along with representative micromagnetic configurations for 
		the reference nanoparticle ($\textsl{\textrm{a}}=\mathrm{150 \; nm}, \; w=36 \; \mathrm{nm}$).}
	\label{fig:diam1Sk}
\end{figure}

In \textbf{Fig. \ref{fig:diam1Sk}} the calculated skyrmion diameter $d_{sk}$ is shown as a function of the external field in the [-0.64, -0.16] T range where $S\approx1$. The $d_{sk}$ depends more or less linearly on $B_{ext}$ 
and a linear regression of the simulation points give $d_{sk}(B_{ext}) =  87.074 B_{ext} + 78.18 $ with a correlation coefficient $R=0.9817879$.  The formed skyrmion at $B_{ext}=-0.16$ T 
has its maximum diameter value close to 62 nm.  Further decrease of the magnetic field causes the gradual decrease of the skyrmion diameter.  Just before its annihilation at  $B_{ext}=$-0.64 T skyrmion attains its minimum  diameter which is  15 nm. It is evident that the  external magnetic 
field plays a dominant role on the actual size of the N\'eel-skyrmion.
\begin{figure}[!h]
	\centering
	\includegraphics[totalheight=5.2in,angle=0]{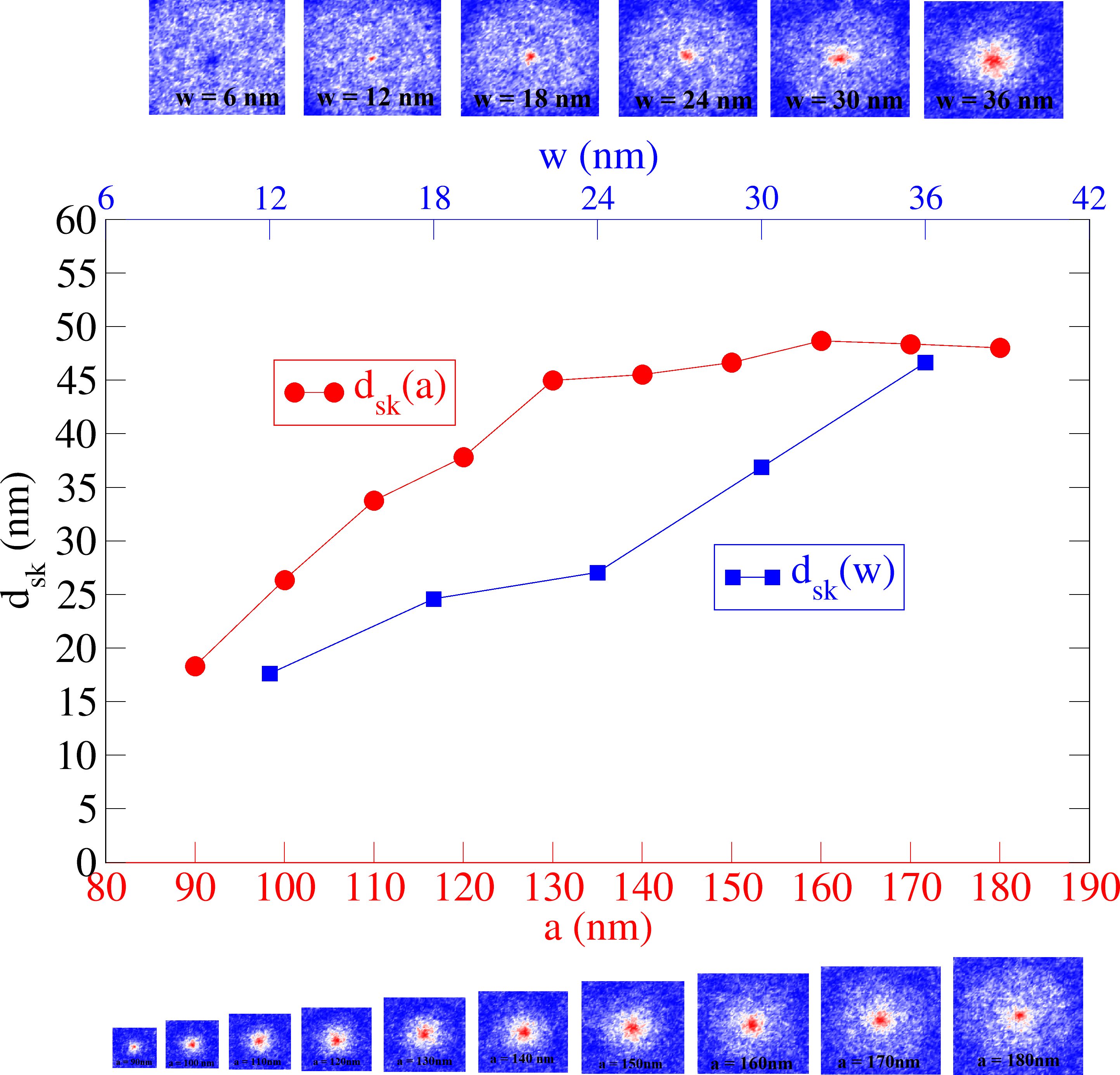} 
	\caption{Skyrmion diameters $d_{sk}$ at  $T=300$ K, $Ku=\mathrm{130.4 \; kJ/m^3}$, $B_{ext}=$-0.4 T as a function of nanoparticle's width $w$ in blue color (side length was set to $\textsl{\textrm{a}}=\mathrm{150 \; nm}$) and as a function of the side length ($\textsl{\textrm{a}}$) in red color (width length was set to  $w=\mathrm{36 \; nm}$).}
	\label{fig:wadiam}
\end{figure}

The effects of the geometrical characteristics of the 
nanoparticle on the formed skyrmion size 
are also studied.  N\'eel skyrmion diameters $d_{sk}$ for  $Ku=\mathrm{130.4 \; kJ/m^3}$, external field value $B_{ext}=\mathrm{-0.4}$ T at $T=300$ K were calculated as functions of nanoparticle's width ($w$) and square's side length ($\textsl{\textrm{a}}$). In \textbf{Fig. \ref{fig:wadiam}}   $d_{sk}(w)$ functional form is shown in blue color with the 
side length set to $\textsl{\textrm{a}}=\mathrm{150 \; nm}$ and 
 $d_{sk}(\textsl{\textrm{a}})$ is presented in red color having constant  
width $w=\mathrm{36 \; nm}$. The increase of $w$ is followed by a monotonic 
increase of $d_{sk}$ values from 17  to 46 nm. 
This monotonic decrease consists of two different linear regions. The first 
one for nanoparticles with width between 12 and 24 nm 
($ d_{sk}(w)=0.78583w + 8.9417$ and $R=0.9639279$). A second linear 
region ($ d_{sk}(w)=1.6342w -12.162$ and $R=0.9999989$) with higher slope is evident for nanoparticles having 24 to 36 nm width. The width value $w=24$ nm represents a critical value for the creation of skyrmions with intermediate sizes having diameters close to 27 nm. 
Regarding the effect of the square side length $\textsl{\textrm{a}}$ of the nanoparticle on the skyrmion diameter also reveals two regimes  
as can be seen in \textbf{Fig. \ref{fig:wadiam}}. The first one is linear and follows the increase of $d_{sk}$ from 15 nm to 45 nm as  
the side length $\textsl{\textrm{a}}$  increases from 90 nm to 150 nm. 
The second one is a plateau regime in which the skyrmion diameter remains 
almost constant at 45 nm for square side lengths up to 180 nm.

\begin{figure}[!t]
	\centering
	\includegraphics[totalheight=5.4in,angle=0]{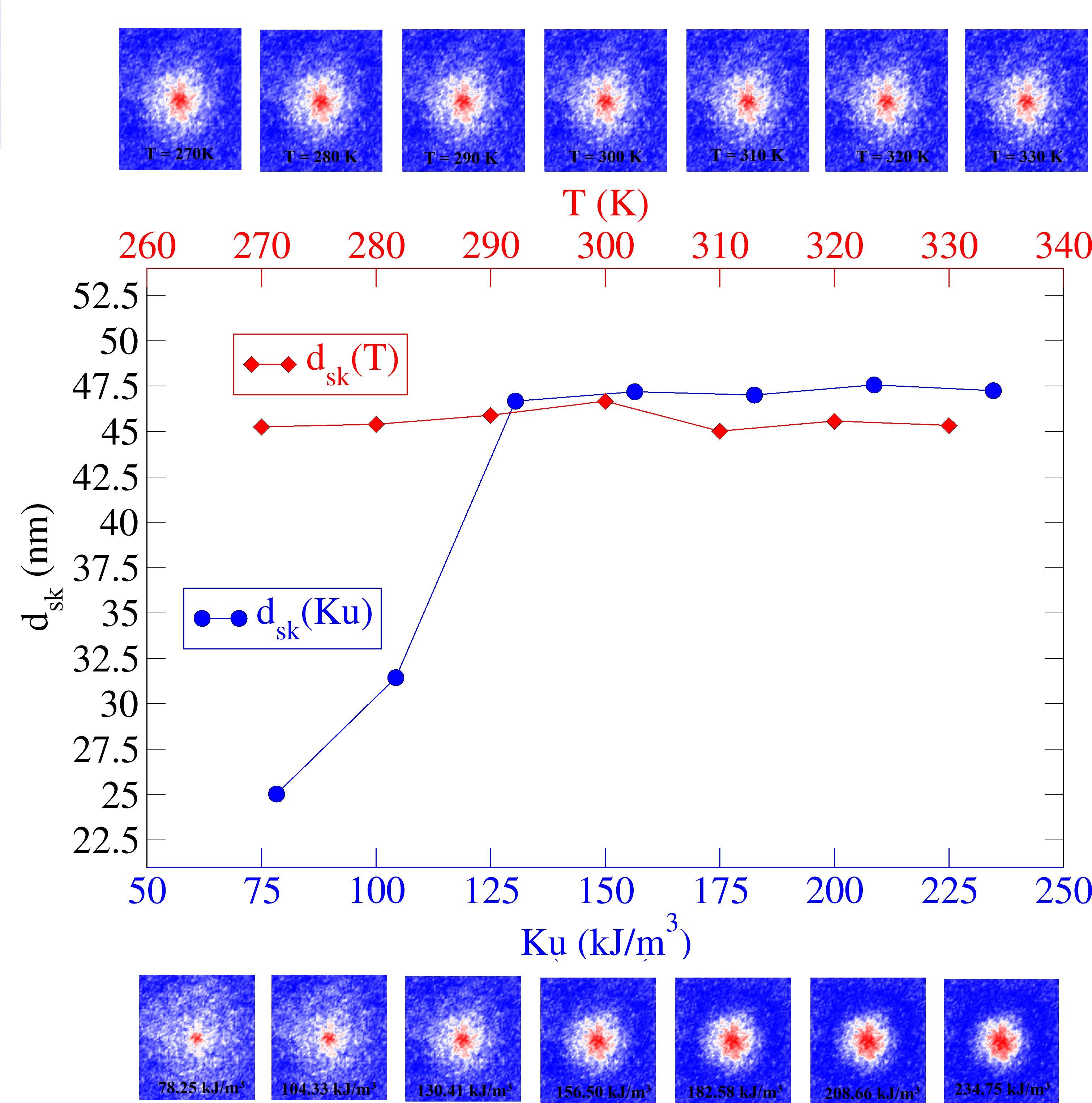} 
	\caption{Skyrmion diameters $d_{sk}$ for magnetic field value $B_{ext}=\mathrm{-0.4 \; T}$ as a function of temperature $T$ (red horizontal axis) and of MCA $Ku$ (blue horizontal axis) at $T = 300$ K.  The reference square-based nanoparticle was used.}
	\label{fig:sizeTandMCA}
\end{figure}

The skyrmion diameter-size has been calculated at different 
MCA values for the reference nanoparticle and the results are given in  \textbf{Fig. \ref{fig:sizeTandMCA}} 
for the specific value of $B_{ext}=-0.4$ T.  At first glance 
two  regimes exist describing the skyrmion's size  dependence on the MCA value. Lower MCA values have a stronger  effect on $d_{sk}$. In particular, as MCA value increases linearly ($d_{sk} = 0.4148Ku -8.8949$) 
from 78.2, to 104.3 and finally to 
130.4 $\mathrm{kJ/m^3}$ the diameter of skyrmion  
increases attaining the values  $d_{sk}\approx$ 25 , 31, 47 nm, respectively.  
Further increase of the MCA
to values up to $Ku=\mathrm{234.7 \; kJ/m^3}$ does not affect the diameter of the skyrmion and a plateau region is evident with the created skyrmion having $d_{sk} \approx 47.5$ nm.    

Finally, the skyrmion diameter for the 
reference nanoparticle and for  $Ku=\mathrm{130.4 \; kJ/m^3}$,  $B_{ext}=\mathrm{-0.4 \;T}$ has been calculated for different 
temperatures ($d_{sk}(T) $) around 300 K as can be seen also 
in \textbf{Fig. \ref{fig:sizeTandMCA}}. It is clear that the temperature does not affect the size of the created skyrmion even for temperatures 30 K above or below the room-temperature. 

\section*{Conclusions}
The skyrmion formation in FePt nanoparticles were 
studied using FD micromagnetic simulations. The adopted 
micromagnetic model takes into account thermal effects 
in the form of a Brownian term in the effective field 
involved in LLG equation and the magnetic material 
properties were reduced to the temperature range of 270 - 330 K. 
MCA and external field were set normal to nanoparticle's surface in all conducted numerical simulations. Magnetic skyrmions 
have been detected during the 
magnetization reversal process for FePt  square-based parallelepiped nanoparticles having different  
side lengths and  widths with MCA value kept constant 
at $Ku=130.4$ kJ/m$^3$ and $T=300$ K.  Computation of the topological 
invariant of skyrmion number $S$ accompanied by 
the visualization of the actual micromagnetic configurations have provided  
detailed quantitative and qualitative information relative to skyrmion 
formation and stabilization. Interesting magnetic structures were 
revealed  for particular external magnetic field 
value ranges. N\'eel type skyrmionic textures were 
detected on the outer surfaces coexisting with Bloch-type 
textures in the bulk of the FePt square-based parallelepiped nanoelements. 

The computed sizes of the created N\'eel skyrmions  showed a 
linear dependence with respect 
to the external field. At different temperatures and for the external magnetic field value of $B_{ext}=-0.4$ T skyrmions can be generated with a 
diameter around 45 nm for the reference nanoparticle having $\textsl{\textrm{a}}=\mathrm{150 \; nm}$. 
Variation of MCA have a significant effect on the skyrmion diameter only for  $Ku \in$ [78.2, 130.4] $\mathrm{kJ/m^3}$  
while for the higher MCA does not affect the diameter which 
attains values close 
to 47.5 nm. The nanoparticles' side length ($\textsl{\textrm{a}}$) 
and width ($w$) affect the skyrmion diameter. The increase of $w$ is followed by an increase of the skyrmion diameter  values from 17 to
46 nm. The side length increase can give birth to skyrmions with diameters close to 50 nm. In conjunction with previous finite element 
studies at 0 K it is clear that magnetic
skyrmions can be produced around 300 K in a wide range of external fields and for nanoparticles with different geometrical characteristic dimensions  
even in the absence of chiral interactions such as Dzyaloshinskii-Moriya for FePt nanoparticles. 

\section*{Acknowledgments}
C. Tyrpenou is supported through the project ''Dioni: Computing Infrastructure for Big-Data Processing and Analysis.'' (MIS No. 5047222) which is implemented under the Action ''Reinforcement of the Research and Innovation Infrastructure'', funded by the Operational Programme ''Competitiveness, Entrepreneurship and Innovation'' (NSRF 2014-2020) and co-financed by Greece and the European Union (European Regional Development Fund). 
 V. D. Stavrou was supported through the Operational Programme
''Human Resources Development, Education and Lifelong Learning'' in the context of the project ''Strengthening Human Resources Research Potential via Doctorate Research'' (MIS-5000432), implemented by the State Scholarships Foundation (IKY) co-financed by Greece and the
European Union (European Social Fund-ESF). We would like to thank
Mr. K. Dimakopoulos for technical support. 

\bibliographystyle{unsrt} 
\bibliography{main}

\end{document}